# Introduction to statistical physics of media processes: Mediaphysics

**Dmitri V. Kuznetsov[†] and Igor Mandel[‡]**

[†]Media Planning Group, 101 Huntington Avenue, Boston, MA 02199, USA
(Affiliated member of Institute of Biochemical Physics RAS)
[‡]Media Planning Group, 195 Broadway, New York, NY 10007, USA
dmitri.kuznetsov@mpg.com, igor.mandel@mpg.com

*Abstract*

Processes of mass communications in complicated social or sociobiological systems such as marketing, economics, politics, animal populations, etc. as a subject for the special scientific discipline – "mediaphysics" – are considered in its relation with sociophysics. A new statistical physics approach to analyze these phenomena is proposed. A keystone of the approach is an analysis of population distribution between two or many alternatives: brands, political affiliations, or opinions. Relative distances between a state of a "person's mind" and the alternatives are measures of propensity to buy (to affiliate, or to have a certain opinion). The distribution of population by those relative distances is time dependent and affected by external (economic, social, marketing, natural) and internal (mean-field influential propagation of opinions, synergy effects, etc.) factors, considered as fields. Specifically, the interaction and opinion-influence field can be generalized to incorporate important elements of Ising-spin based sociophysical models and kinetic-equation ones. The distributions were described by a Schrödinger-type equation in terms of Green's functions. The developed approach has been applied to a real mass-media efficiency problem for a large company and generally demonstrated very good results despite low initial correlations of factors and the target variable.

## *1. Introduction*

The object of the present paper is a methodology of application of statistical physics to the complicated phenomena of mass communications. These phenomena take place practically in all social and sociobiological systems and have been already widely studied under different names and with different approaches. In our opinion, they play such a significant role, that they are worth not only a devoted field of studies, but also the application of dedicated methodologies. Thus this article outlines a field that we called "mediaphysics" and proposes a general physical model, capable of serving that field. Mediaphysics may be determined as a *part of sociophysics, studying processes of mass communications in social and sociobiological systems;* it has its own subject and operates with data of certain types.

**1. Subject - mass communications**. Communications between units (people or animals) in some important sense have similar nature in social and sociobiological systems (Wilson, 2000). It was studied earlier from different perspectives; however the importance of that field encourages some isolation of the subject. The importance of the subject was already realized, for example, in an excellent recent study of social insects behavior (Bonabeau, Dorigo, Theraulaz, 1999), which is based entirely on a concept of communications mechanisms.

**2. Data - observed and unobserved.** Let's consider two datasets: (a) 20 weekly observations of J. Smith's Coca-Cola purchases, and (b) 20 weekly observations of Coca-Cola sales in the USA. Traditional statistics will treat those sets identically. However, they are qualitatively different. In case (a) the level of purchase is explained by individual behavior of J. Smith, his habits, income, etc. In case (b) the same explanation is applied too (all individuals, buying soda in the USA, have motivations, similar to J. Smith's one); but on a top of that, there is a *structure* of these individuals by their *distribution* in readiness and willingness to buy soda in general and Coca-Cola specifically. The last distribution is changing over time and that is something very important which does not exist in case (a). The sales value for the USA is in fact a result of millions of individual actions, which appears mostly in an unobserved form. The observations are just the tip of the iceberg and reflected in the sales volume. Mediaphysics can work with both observed and unobserved data in a single model.

Traditionally, social and biological sciences used statistics (not statistical physics) to describe the above mentioned phenomena, ignoring the deep differences between two data types and using just observed ones. We propose to analyze systems like (b) using developed techniques of statistical physics. It is an appropriate tool due to the deep similarities between behaviors of particles or molecules *en mass* (typical unobserved units), and processes described above. However, it does not mean that mediaphysics should employ only approaches like the proposed one.

The placement of mediaphysics into sociophysical realm needs some comments, to describe why it seems necessary, and to provide historical and linguistic connotations. Here we briefly outline some general points.

**Sociophysics** has acquired growing popularity for the last few decades, but in our opinion its field is not clearly determined yet. It should be logically and historically rooted from the first ideas of socio-physical analogies by Vico and de Condorcet through the famous "Physique sociale" by A. Quetelet (1835; second edition in 1869). In modern times the term was reestablished in (Galam, Gefen and Shapir, 1982), without reference to the mentioned "fathers-founders". Instead, in 2004, S. Galam makes an elegant statement: "I feel to claim paternity over Sociophysics…" though mitigated with reasonable concern that "…paternity always contains a little bit of a doubt, how to perform an ADN check" (Galam, 2004). Well, we are not DNA test specialists, but able to testify that if our translation of Physique Sociale as Social Physics, and further revolutionary abbreviation of the latter to Sociophysics is correct, then the worst of Galam's suspicions about paternity (or even "great-grand-childhoodedness") may be, indeed, right. Also, some works without Sociophysics name (but about the same subject) appeared earlier (Weidlich 1971; Gallen and Shapiro, 1974; see also reviews in the books by Schweitzer, 2003, and Weidlich 2000).

A similar uncertainty exists about Sociophysics subject. In (Arnopoulos, 1993 and 2005), the only modern books we know under that title, the sociophysics has a super-universal meaning and embraces both nature and social life within entire paradigm *(universal definition)*. The latest reviews (Stauffer, 2003, 2005), do not provide exact definitions, but actually describe a much more narrow field, limited mainly (but not only) by computer simulations of social phenomena based on (supposedly, but not always in practice) physical principles *(simulation definition)*. Characteristically, the authors of those "two schools" do not cite each other, although the first edition of Arnopolous's book took place in 1993. For our purposes, we may tentatively define sociophysics *as a science about the application of physics in social sphere* (*broad definition*).

Another related field with established name, **econophysics,** falls into that broad definition. It is mostly dedicated to applications of statistical physics to stock-options pricing and portfolio optimization (Mantegna and Stanley, 2000; McCauley, 2004). There is also smaller number of studies in the more traditional economical areas (Econophysics of Wealth Distributions, 2005).

In the most popular simulation sense, sociophysics typically considers a society represented as a set of interrelated and interacted nodes (people) on lattice (Ising model and its generalizations) or a random graph, subject to further computer simulation of its dynamics (see also review: Sznajd-Weron, 2005). In models like that, a certain number of randomly selected agents, either convince their neighbors of their opinion, or adjust their own opinion, or adjust both their and neighbors opinions. It could be called **Social Simulation**, where physics is not necessarily involved. The corresponding analytical approaches are less developed. However, many interesting analytical studies were published during last few decades (see, for instance, Weidlich, 1971; Schweitzer, 2003; Slanina and Lavicka, 2004; Wu and Huberman, 2004; etc.). A big plus of the simulation approach is its simplicity for formulation of a simulated scenario, and hence the number of possible models is limited only by researcher's curiosity. A lot of interesting results were already obtained in that direction. However, this approach has some intrinsic problems.

Simulations have significant limits in numbers of objects and involved factors. And these restrictions are, indeed, observed in majority of publications that makes them *far from measured reality*, while capturing some directional and qualitative observations. For the history of sociophysics simulation (at least from about 50 papers published last five years we reviewed) just a few examples of real applications were presented. Typically, sociophysics simulation models do not use real data to estimate models parameters. Such freedom from reality allows, on one hand, checking exotic hypotheses, but on the other hand, leaves too much room for arbitrary decisions. One of the most cited applications of the sociophysics simulations is a modeling of election results in Brazil, that was called "…at present the strongest validation of the Sznajd Model" since its introduction in 2000 (Fortunato, 2005). However, while not disputing the results of that study, it is doubtful, that real life election results depend only on a positive interaction of people; and that advertising, negative word-of-mouth effects, and the political programs of candidates (factors, which were not considered) do not affect the results.

Modern causal theory (Pearl, 2000) applies very sophisticated methods just to separate "*causation vs. correlation*", a problem that in physics usually stands in a very different manner (see, for instance, Arnopoulos, 2005) and usually doesn't stand in simulations at all. The danger is that impressive results in an idealized world may be even *directionally wrong* in real application. For instance, if only interactions between people can explain the election results, as in the cited example above, it can be easily imagined that if one would apply only advertising he could get the same results. It may happen because advertising is a cause of neighbors convincing (people are convinced not because they are neighbors, listening each other, but because they also watched the same TV show yesterday). This effect is widely known as multi-colinearity (one "cause" is to be explained by another correlated with the first one) in statistics, but not in statistical physics. So, there is a methodological gap between these two ways of looking at the same social phenomena. The current sociophysics-simulation view is extremely useful showing things from very different prospective and challenging the traditional statistical techniques, but does not fill the gap by itself.

The introduced mediaphysics overlaps with discussed above sociophysical interpretations and statistics, but focuses only on communications and thus belongs to the sociophysics field inside its universal and broad definitions. The mediaphysics orients to real-life data, what currently is not typical for sociophysics simulations. It deals with both observed and unobserved data unlike traditional statistical approaches. Plus, mediaphysics is associated with two meanings of the term "media", both of which are relevant to the approach: media as an environment in which mass processes of communications are taking place; and, in a form of "mass-media", as an advertising (or other messages) spreading through mass communication channels, which itself is a very important topic.

The article is organized as follows. Section 2 contains a motivating example and defines new terms (in spite of its simplistic style, it introduces important concepts used thereafter, especially concept of mindset's space and motion in there); in part 3 we describe main formalism, based on Green's functions and Schrödinger-type equation; section 4 demonstrates how this approach is related with two important modern statistical techniques; then in 5 we discuss model implementation to the real data; and part 6 presents conclusion remarks.

## *2. Competitive fishing and distributions in a space of persons' mindsets*

In the modern world people are subject to hundreds of activities intended to attract customers, voters, or followers. It creates a strong competitive environment and can be, roughly but quite reasonably, demonstrated in terms of *a competitive fishing* illustrated in Fig. 1. This analogy reveals many features, useful for further formalization.

Fishing success depends on your and your competitor's positioning and *the amount of fish that are swimming around you in the current moment of time*. This amount is proportional to the volume of yours fishing area and the corresponding fish density there (Fig. 2). In the meantime, the present fishing conditions and its dynamics are resulted from natural history, and caused by your and your competitor's previous fishing activities. Thus, the current density in a fishing area contributes to the current catch of the area's fisherman through his current baiting activity but, simultaneously, the current baiting activity can change the properties of the fishing area for the future. There are two main factors in the fishing dynamics.

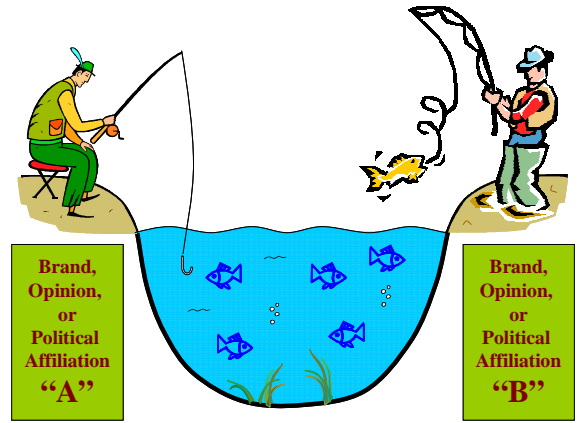

**Figure 1. Competitive fishing and underwater world**

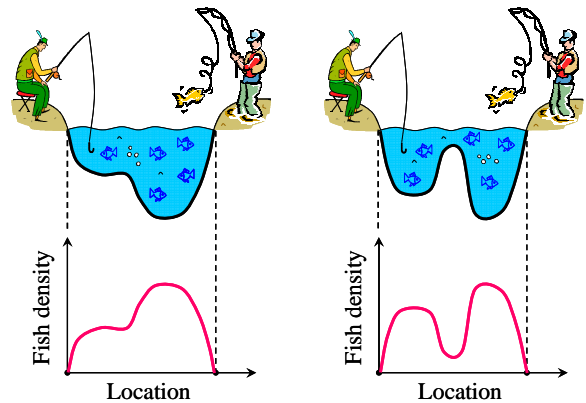

**Figure 2. Different density shapes**

(1) **Non-competitive factor**: some fish can swim *in or out* of the fishing areas, increasing or decreasing *total* future catch. The biggest portion of the fish population is usually swimming outside the fishing areas and they cannot be caught at that moment neither by you nor competitor, but under certain circumstances they can move in.

(2) **Competitive factor**: fish can swim *between* the fishing areas, increasing or decreasing *your* future catch. In Fig. 2a and 2b we demonstrate fishing in different environments in terms of volumes of fishing areas and density distributions of the fish population. A barrier between fishing areas in Fig. 2b can complicate or even prevent penetrations between the neighboring fishing areas. The last case reflects two almost non-overlapping fish markets with almost no *short-term competitive effects*; this state can take place until barrier destruction (*long-term competitive effects*) leading to significant increase in possibility to relocate and attract catch from competitors.

Thus, analysis of density distributions and their transformations is a tool to deal not only with short-term but also long-term effects from external and competitive factors. Let us consider the analogy a little deeper to establish dynamics of fish-population distribution based on individual fish behavior.

**Single-fish "free" swimming.** A single fish can be driven by food and some other attractors and/or repellents. We assume that if there are no attractors, repellents, space boundaries and barriers around, it moves just randomly in any direction searching for food or other attractors (we will call it a "free fish"). Naturally, each fish cannot move too far from its previous position in any given time interval: the higher distance, the lower probability to be there (Fig. 3). It means that probability distribution of the next position depends on its previous one and therefore this is the Markovian process. On the space scales much larger than fish size, this distribution often can be reasonably well described by normal curve (Gaussian behavior with direct Brownian-motion analogy), but generally speaking it is not always true (for example, for a sort of fish which remembers its orientation in space for a long period of time). For any symmetric distribution of next-step location probability, one of the most important physical characteristics is the root-mean-square displacement $a$ (see Fig. 3 and comments in 3.1), which is a measure of fish's *"flexagility"*. This term, a combination of a mind's flexibility and a person's agility, is coined to replace more technical "displacement" in social applications, or more precisely, to reflect causes of the human ability to be displaced far enough from the previous state. A highly flexible person often changes directions of the movement, an agile person moves fast in a given direction. Thus, a low flexibility combined with a high agility always provides a high displacement, while a high flexibility together with a low agility guarantees a low displacement. Two other combinations lead to different results depending on the values of these parameters. A person is flexagile if he/she changes states of mind (opinions, preferences, etc.) easily and fast in the absence of external forces.

**Fish types.** It is natural to define that two or more fishes are of the same type (i.e. the group is homogeneous) if they all have the same next-moment position distribution (Fig. 3) with respect to their corresponding current states (current states are different for each fish). Thus, all the fish from one homogeneous group are characterized by the same root-mean-square displacement $a$. By this definition, some fishes can be different biologically but be the same in their behavior and, therefore, belong to the same type here. According to the definition, different fish types have different next-moment position distributions and different $a_i$ values, where $i$ stands for the fish type. For a large fish population, $a(i)$ can be distributed continuously with its own probability distribution.

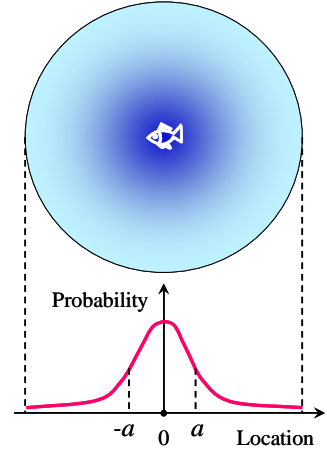

**Figure 3. Probability to move to some distance from a given location**

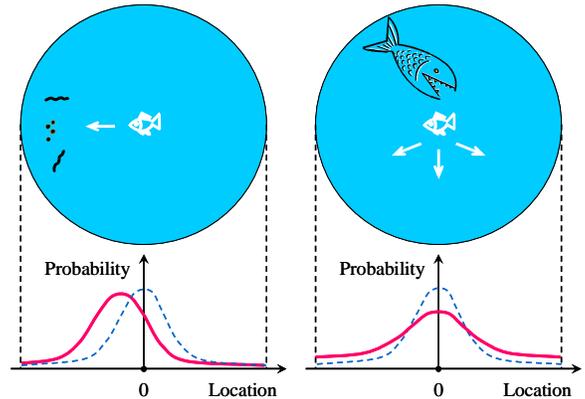

**Figure 4. (a) Distribution is shifted by location, (b) distribution is flattered**

**Single-fish swimming with attractors, repellents, barriers and boundaries.** In reality fish do not always move randomly. It is attracted to some locations (by food, bait, etc.) and repulsed from others (by something dangerous). These factors can change next-position probability distribution for each fish by giving a priority to some swimming direction (Fig. 4a vs. "free" swimming in Fig. 3) and/or transforming the distribution to a more flat form (Fig. 4b). The process is still Markovian. However, the movement is now described not only by "free fish" root-mean-square displacement $a$ (restriction for too far moves), but also its aspirations (forces), which are caused by different stimulus (fields). Mathematical aspects of these effects will be described below in section 3. Here, we would like to mention that forces (or fields as their originators) applied to fishes can have very a different nature and characteristics, and be caused also by "internal" fish factors, such as interactions between fishes: for instance, it could be a food information transduction (influence) inside fish community. Boundaries and/or obstacles can disturb the fish behavior as well. For example, a "free fish" in a closed area in a long enough period of time can be found everywhere in that area with equal probability.

**Dynamics of fish-population density distribution.** Each fish has a single position at each moment of time, contributing to population distribution over the available space. Then, each fish from that distribution moves and has own next-moment position probability distribution (Fig. 3 and 4) in respect to its previous position (which is known). As we have mentioned, these individual distributions can be different for different fishes and undergoes an influence of attractors, repellents and internal interactions. The sum of individual next-moment position distributions for all fishes, constructed around their corresponding individual previous positions is a new (next-step) population distribution. In a continuous limit, the sum has to be replaced by integral over available space of the individual next-step distributions with previous population position distribution taken as a weight factor. This procedure can be repeated many times to describe dynamics of the fish population, based on the individual fish behaviors, and accumulating all available factors, including previous baiting activities, both their own and competitor's ones. (See mathematical description in 3.1.)

**Back to social life.** After we have exploited the fishing analogy deeply enough, we can construct a frame for social life analysis, where all considered aspects will naturally fall into. We will discuss it using an example of marketing activity, keeping obvious links with voting and religious or political affiliations, and still referring to fish example when appropriate. Instead of fishes and their distributions in water, we will consider personal mindsets in a space between two (Fig. 5) or many choices (Fig. 6), and their distributions for a human population. Changes in personal mindsets (stimulated by many factors including advertising activities) can lead to transformations of distributions over time. Here, we are using several definitions that are worthy to emphasize especially.

- **Units** are individuals, who make some decisions: fish, decided where and what to eat; people, decided where and what to buy (what party to join, what religion to follow, etc.).
- **Space**, **in which the units move**. It is a river's basin for fish; a virtual space of "willingness to buy" (or "propensity to join or believe") within a person's mindset for people.

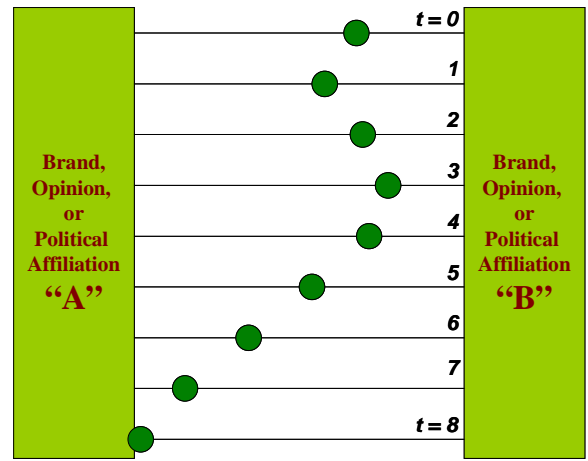

**Figure 5. Random walk between two opinions in one person's mind (floating mindsets)**

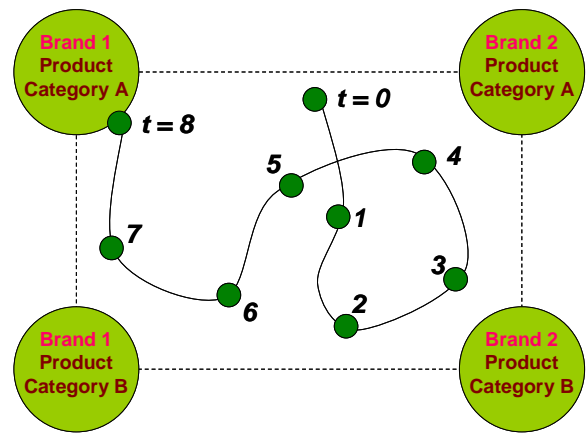

**Figure 6. Random walk between many opinions**

- **Dimensionality of this space.** It depends on model. If we consider just two alternatives, say, the willingness to buy products of a specific brand name vs. any other brands, then the dimensionality is one, because with two choices only a unit's position along the line connecting two alternatives matters. Extra dimensions can arise from explicit multi-brands competition, competition between categories of products inside single brands, and some other real-life complexities.
- **Special dimension of the space may be called "demand"**, which reflects a desire (supported by financial ability) to buy any product in a specific category (for instance, any nonalcoholic beverages).
- **Next-moment position distribution for a "free" single unit** can be usually characterized by the root-mean-square displacement $a$ that can be different for different unit types. The displacement $a$ is a measure of personal flexagility. The higher the flexagility, the higher the probability that a person changes significantly his/her opinion within the next moment, regardless of previous state.

- **Population distribution** is a normalized sum of all single-unit distributions, calculated using previous population distribution as a starting point for each unit.
- **Catch** is a number (or portion) of units absorbed on (caught by or bought) an alternative (brand, opinion, fisherman's baiting hook, etc.). This is an observable or measurable value, which reflects only one point or a specific part from the entire population distribution.
- **Observed and unobserved parts of the population.** Usually many parts of the population distribution are not directly observed, but contribute to future catches due to unit motions and, therefore, the knowledge about an entire distribution is very important. Thus, if a fisherman knows, that a huge fish shoal is approaching (the distribution is skewed to his shore, but not observed yet), he will not pack his stuff and go home. The same is true in marketing. The mediaphysical approach can analyze and forecast the hidden distribution forms. By extra measurements (e.g. surveys), some hidden parts of a distribution can become observed, providing an additional knowledge about the distribution shape. Thus, these measurements are very important and desirable for the understanding of hidden processes and model tuning.
- **Fields** are considered as originators of forces applied to units and cause their trends to move in specific directions. In marketing, those are reduced factors of advertising, economical situation, seasonality, influential effects ("mouth-to-mouth" information that is very hard to count in traditional statistical approaches, but what is a hot topic in sociophysics), price, and so on. Fields are functions of the corresponding variables that are described in 3.3. The field approach is based on principles, which, being logically developed, lead to important statistical concepts postulated in the "adstock" paradigm for marketing with prolonged lag effects and the "yield" analysis with individual-effect estimations in each data point (see chapter 4).
- **Homogeneity or heterogeneity** of the population is taken into account by considering the same or different next-moment position distributions for a "free" single unit (i.e. the same or different the root-mean-square displacements $a_i$). The organic way to incorporate population heterogeneity is another serious advantage of the proposed approach from traditional methods.
- **Initial conditions for population distribution** is starting point for analysis of the system dynamics. They are based on substantive knowledge of the system. For marketing, initial population distribution has to be coordinated with the market share, which is usually known.

All these components are used to construct a formal description of the mediaphysical model.

## 3. Physical approach to the problem

Here we describe a mathematic (or statistical physics) technique to analyze population distribution between two or many alternatives. Relative distances between a person's mindset and the alternatives are measures of person's propensity to buy, see Fig. 5 and 6. The state of a person's mind depends on its state at the previous moment of time and also is subject to the action of many factors of different originations, including external (economic, social, marketing and so on) and internal ones (influential propagation of opinions, synergy effects, etc.).

As it has been mentioned above, the fundamental feature of the considering system is a Markov chain probability of each unit to be within certain distance from previous state at any given time. This probability is determined by the flexagility in a mental space. It makes the process very similar to the Brownian motion under action of external fields. The corresponding mathematical approach is well developed in statistical physics and based on Green's functions.

### 3.1. Green's functions

The Green's function $G_t(\vec{q}_t, \vec{q}_0)$ is the conditional probability that the state of person's mind (or the Brownian particle) at time $t$ is placed at the point $\vec{q}_t$ provided the initial state was at the point $\vec{q}_0$. The $\vec{q}_t$ can be one or multi-dimensional *"generalized coordinate"* of the state. In particular, it can include not only "real coordinate" but also, for instance, a velocity. The Green's function $G_{t+1}(\vec{q}_{t+1}, \vec{q}_0)$ has the recurrence relation

$$G_{t+1}(\vec{q}_{t+1}, \vec{q}_0) = \int G_1(\vec{q}_{t+1}, \vec{q}_t) \cdot G_t(\vec{q}_t, \vec{q}_0) d\vec{q}_t \quad (1)$$

or

$$G_{t+1} = \hat{Q} G_t, \quad (2)$$

where $\hat{Q}$ is the transfer operator written as

$$\hat{Q} = \exp[-W_{t+1}(\vec{q})] \cdot \hat{g}. \quad (3)$$

The $W_t(\vec{q})$ stands for a field (in energetic temperature units for Brownian particle) applied to units in position $\vec{q}$ at time $t$, and $\hat{g}$ is the unit-connectivity operator which, if applied to an arbitrary function $\psi(\vec{q})$, can be written as

$$\hat{g}\psi(\vec{q}) = \int g(\vec{q}, \vec{q}')\psi(\vec{q}')d\vec{q}'. \quad (4)$$

The function $g(\vec{q}, \vec{q}')$ describes the connection of neighboring states during time in terms of the Markov chain conditional probability in the absence of any other fields and, therefore, it is a distribution of personal flexagility. In the absence of fields (i.e. where $W_t(\vec{q}) \equiv 0$), equation (3) is reduced to $\hat{Q} = \hat{g}$ that has to be for "free" units.

For the simplest Gaussian model in the D-dimensional position space $\vec{z}$ (assuming $\vec{q} \equiv \vec{z}$), i.e. where

$$g(\vec{z}, \vec{z}') = \left(\frac{D}{2\pi a^2}\right)^{D/2} \exp\left[-\frac{D(\vec{z} - \vec{z}')^2}{2a^2}\right] \quad (5)$$

follows to the normal distribution like in Fig. 3, in leading terms (see, for example, book: Grosberg and Khokhlov, 1994)

$$\hat{g} \approx 1 + \frac{a^2}{2D}\Delta_{\vec{z}} \quad . \quad (6)$$

Here $\Delta_{\vec{z}}$ is the Laplacian operator in the position space $\vec{z}$, and $a^2$ is the mean-square distance (displacement) between two neighboring states during time in the absence of other fields:

$$a^2 = \int \vec{z}^{\,2} g(\vec{z}, 0) d\vec{z} \quad . \quad (7)$$

The operator $\hat{g}$ can be reduced from general integral form (4) to a differential one, like in (6), not only for the Gaussian case but also for many other different models (see, for instance, the case of an additional orientation memory in Grosberg, 1979; Kuznetsov and Sung, 1997).

We should emphasize that equations (3) – (6) are valid for a smooth variation of the external field $W_t(\vec{q})$ on the time scale of each model. The reason is that for these cases we can consider the interval between two neighboring states as unperturbed by external fields and identify (2) as the Chapman–Kolmogorov equation for transitional probabilities.

In continuous limit (for a large number of time intervals $t$), we can rewrite (2) as

$$-\frac{\partial G_t}{\partial t} = (1 - \hat{Q})G_t \quad . \quad (8)$$

For smooth fields equations (3), (6) and (8) lead to the Schrödinger-type equation:

$$\frac{\partial G_t}{\partial t} = -W_t \cdot G_t + A \cdot \Delta_{\vec{z}} G_t \quad , \quad (9)$$

where $A \equiv a^2/2D$ is a constant. Although (9) has the form of time-dependent Schrödinger equation, it does not include any quantum effects here.

Let us assume that the total population in the considered system is $N$. Generally speaking, $N(t)$ can be time-dependent, because of a possibility for population to grow or shrink. This value is incorporated through the normalization condition:

$$\int G_t(\vec{q}_t, \vec{q}_0) d\vec{q}_t = N(t) \quad . \quad (10)$$

For a specific problem, equation (9) can be significantly simplified. Thus, for 1-dimensional positional space $\vec{z} = z$ the Laplacian operator is reduced to $\Delta_{\vec{z}} = \frac{d^2}{dz^2}$.

Moreover, for numeric calculations it is convenient to come back to discrete form of equations (assuming that $\frac{\partial G_t}{\partial t} \approx G_{t+1} - G_t$), then equations (2) and (9) are transformed to

$$G_t = (1 - W_t) \cdot G_{t-1} + A \cdot \Delta_{\vec{z}} G_{t-1} \quad . \quad (11)$$

### 3.2. One-dimensional space for two alternatives

Let us consider a simple case of two brands like in Fig. 5. Space between the brands $0 < z < 1$ is filled by a large number of persons. Each person is associated with a point $z$ in that space. The distances $z$ and $1 - z$ between the point and brands correspond to a person's mindset with respect to each brand respectively. The shorter the distance, the stronger the mindset is that this person has to that brand. There is a distribution of population between brands for each moment of time. Assuming an initial form of the distribution $G_{t_0}(z)$ at time $t = t_0$ we can calculate step by step corresponding distributions $G_t(z)$ for the moments $t = t_0 + 1$, $t_0 + 2$ and so on. Here, for 1D calculations,

$$G_t = (1 - W_t) \cdot G_{t-1} + A \cdot \frac{d^2 G_{t-1}}{dz^2} \quad . \quad (12)$$

The last equation is valid for $0 < z < 1$. To calculate $G_t(z)$ at $z = 0$ and $z = 1$, we have to apply boundary conditions, which depend on considering a specific problem and are selected from a "physical meaning" of the boundaries. For instance, for non-absorption case (the only store is closed, or the fisherman is sleeping and does not make a catch): $G_t(0) = 0$; for sales with in-stock daily limits that are lower than demand: $\frac{\partial G_t(0)}{\partial t} = 0$; for uniform near-attractor area: $\left.\frac{\partial G_t(z)}{\partial z}\right|_{z=0} = 0$, etc.

### 3.3. Fields

The total field $W$ is affected by many factors including economic, social, marketing, interpersonal opinion influence and others. It can be presented using different level complexity corresponding to specific objectives. We plan to describe it in details in our next paper. Here, we only demonstrate the basic ideas behind the definition of fields, and, therefore, we present $W$ in its simplest form. In this form the total field can be presented as a sum of the following main pieces:

$$W(\vec{z},t) = W_0(\vec{z},t) + W_C(\vec{z},t) \\ + W_F(\vec{z},t) + W_I(\vec{z},t) - \overline{W}(t) \quad (13)$$

where $W_0(\vec{z},t)$ and $W_C(\vec{z},t)$ are the field contributions from both own and competitor's advertising activities, respectively; $W_F(\vec{z},t)$ stands for the contributions from general (non-advertising) factors, like economic (for instance, Dow Jones indexes and average national prices for a product category) and social ones; $W_I(\vec{z},t)$ is the influential part, which is based on interpersonal relations and opinion exchange; and $\overline{W}(t)$ is the uniform position-undependable term that added because of the following reasons. The forces, which applied to units, are determined by relative changes of fields over considering position space. Therefore, we can add or subtract a constant field (constant over the space but maybe different for each time moment $t$) without a result perturbation. The most reasonable way is to make the space-average of $W(\vec{z},t)$ equals zero, i.e. $\int_V W(\vec{z},t)d\vec{z} = 0$, where $V = \int_V d\vec{z}$ is the considering space volume. Then, $\overline{W}(t)$ is defined as

$$\overline{W}(t) = \frac{1}{V}\int_V \begin{bmatrix} W_0(\vec{z},t) + W_C(\vec{z},t) \\ + W_{GF}(\vec{z},t) + W_I(\vec{z},t) \end{bmatrix} d\vec{z}, \quad (14)$$

Keep considering a simple form of fields, $W_0(\vec{z},t)$ and $W_C(\vec{z},t)$ can be characterized directly in terms of expenses in different advertising channels. If we have $n_0$ channels of own advertising and $n_C$ channels of competitor's ones, then for 1D system with $\vec{z} = z$,

$$W_0(\vec{z},t) = z^{\nu_0}\beta_0\left[1 + \sum_{i=1}^{n_0} B_{0i}b_{0i}(t)\right] \quad (15)$$

and

$$W_C(\vec{z},t) = (1-z)^{\nu_C}\beta_C\left[1 + \sum_{i=1}^{n_C} B_{Ci}b_{Ci}(t)\right], \quad (16)$$

where the powers $\nu_0$ and $\nu_c$ characterize spatial dependences of the fields with respect to the corresponding attraction centers at $z=0$ and $z=1$; $b_{0i}$ and $b_{Ci}$ are own and competitor's advertising expenses in channel $i$; $B_{0i}$ and $B_{Ci}$ are corresponding weight factors to be estimated; and, finally, $\beta_0$ and $\beta_C$ stand for total scaling factors. The general-factor fields can be written in a similar form:

$$W_F(\vec{z},t) = \left|\frac{1}{2} - z\right|^{\nu_F}\beta_F\left[1 + \sum_{i=1}^{n_F} B_{Fi}b_{Fi}(t)\right], \quad (17)$$

where $i$ stands for type of the corresponding factor $b_{Fi}$ and $n_F$ is the total number of accounting general factors. The scaling factors $\beta_0$, $\beta_C$ and $\beta_F$ are introduced here (unlike the usual presentation in regression models) to have the weight factors $B_{0i}$, $B_{Ci}$ and $B_{Fi}$ as the measures of relative advertising-channel effectiveness, which are more treatable than the corresponding productions $\beta_0 B_{0i}$, $\beta_C B_{Ci}$ and $\beta_F B_{Fi}$. In last equation the attraction center is placed at $z=1/2$, which is somehow arbitrary, but based on the reasonable assumption, that general factors affect people regardless on their proximity to the "real" (own and competitor's) attraction centers. In general, in the people-mindset spaces with dimensionality more or equal two, the array of points of equal distances from different attractors is interesting subject, because closeness to one node does not mechanically mean remoteness from another (for instance, a dualistic distance concept was proposed by Mandel, 1979), which we are going to consider in another article.

Influential part of the fields deserves a special consideration. As we have mentioned above, there exist many attempts to analyze influential factors in opinion propagation in sociophysics. Their majority is based on Ising-type models, which assume sophisticated person-to-person interaction principles. It is clear that consideration of each person individually is not always acceptable, especially for large populations. Here, we propose a different approach that is used in statistical physics to describe volume interactions between particles. This is, so-called, self-consistent mean-field approach, where interactions (or influences here) are determined by density distributions of units over a space of positions (or opinions, as described above). In this paper, we reproduce this approach in the second (pair or unit-to-unit) virial approximation:

$$W_I = \beta_I G(z,t), \quad (18)$$

where $G(z,t)$ is the introduced above Green's function describing a unit density distribution at time $t$, and $\beta_I$ is the scaling factor.

## 4. Links to modern statistical techniques

Among consequences of the basic principles of the proposed mediaphysical approach, we can recognize many concepts postulated in modern statistics. On one hand, this makes the approach naturally related with modern statistics. On the other hand, the links between the mediaphysical approach and some of the concepts make clearer not only the origination of the concepts but also their positioning in a whole picture of analyzed phenomena.

Here we demonstrate the straightforward links of the proposed approach with (a) lag concepts and their modifications including trend analysis, Box-Jenkins models or adstock ones in marketing (Broadbent, 1997; Mandel and Hauser, 2005), and (b) random coefficients mixed models (Demidenko and Mandel, 2005). Both concepts, in fact, are generalizations of large classes of models in time series analysis and regression and very important for multiple applications. To start, let us to remind briefly the forms of basic equations in each of the concepts.

**The regression with lags concept** usually declares the following dependencies of a target variable (say, own sales) $S(t)$ on a set of $m$ factors

$$\vec{b}(t) \equiv \{b_1(t), b_2(t), ..., b_m(t)\} \tag{19}$$

for each moment $t$:

$$S(t) = \theta(t) + \sum_{i=0}^{i_0} \left( \vec{C}_i \cdot \vec{b}(t-i) \right), \tag{20}$$

where $\vec{C}_i \equiv \{C_{1i}, C_{2i}, ..., C_{mi}\}$ are coefficients that are different for each lag $i = 0, 1, ..., i_0$ and $\theta(t)$ is the time-dependable baseline. The coefficients $\vec{C}_i$ can be interrelated and obeyed a specific function of $i$ and a couple of parameters (for example, two-parameter gamma function for adstock models in Mandel and Hauser, 2005). The baseline can include an explicit trend:

$$\theta(t) = \theta(t_0) + (t - t_0) \left. \frac{\partial \theta(t)}{\partial t} \right|_{t_0}. \tag{21}$$

**The random coefficients mixed models** are intended to estimates coefficients (yields) on each data point and can be presented as

$$S(t) = \theta(t) + [\vec{C}^0 + \delta\vec{C}(t)] \cdot \vec{b}(t), \tag{22}$$

if no lag effects, or in a form similar to (20) with lag effects but with fluctuating coefficients for each lag

$$\vec{C}_i(t) = \vec{C}_i^0 + \delta\vec{C}_i(t), \tag{23}$$

i.e.

$$S(t) = \theta(t) + \sum_{i=0}^{i_0} \left( [\vec{C}_i^0 + \delta\vec{C}_i(t)] \cdot \vec{b}(t-i) \right). \tag{24}$$

Here, $\vec{C}_i^0$ is a constant value for each lag $i$ and $\delta\vec{C}_i(t)$ is the fluctuating part of the factor's coefficient. The last equation is a generalization that includes both concepts of regression with lags and random coefficient mixed models.

**The links of mediaphysics with above concepts** can be established from equation (11) and the relation between the target (sales) value $S(t)$ (below in this paragraph we will write it as $S_t$) and the Green's function at the absorption (zero) point $(G_t)_0$:

$$S_t = \lambda \cdot (G_t)_0, \tag{25}$$

where the sales-to-Green's-function scaling factor $\lambda$ is the same for a whole space distribution (in particular, it means that $\lambda$ is the same for both own and competitor's sales). Then,

$$S_t = (1 - W_t^0) \cdot S_{t-1} + A\lambda \cdot (\Delta_{\vec{z}} G_{t-1})_0, \tag{26}$$

where $W_t^0$ is the field at the absorption point at time $t$. Using the recurrence relation (11) with (26) $i_0$ times, we can write

$$S_t = S_{t-1-i_0} \prod_{i=0}^{i_0} (1 - W_{t-i}^0)$$
$$+ A\lambda \cdot \sum_{i=0}^{i_0} \left\{ \left[ \prod_{j=1}^{i} (1 - W_{t-j}^0) \right] (\Delta_{\vec{z}} G_{t-1-i})_0 \right\} \tag{27}$$

Here, $S_{t-1-i_0}$ and $G_{t-1-i_0}$ can be considered as initial sales and population density distribution, respectively.

The fields that composed by factors (19) are explicitly included in (15) while other implicit ones included in other terms of equation (13), therefore we can write

$$W_t^0 = W_t^* + \widetilde{W}_t, \tag{28}$$

where the term with factors (19) is

$$W_t^* = -\vec{B}^* \cdot \vec{b}(t). \tag{29}$$

Thus, for smooth fields the production

$$\prod_{i=0}^{i_0} (1 - W_{t-i}^0) \approx \left( 1 - \sum_{i=0}^{i_0} \frac{W_{t-i}^*}{1 - \widetilde{W}_{t-i}} \right) \cdot \prod_{i=0}^{i_0} (1 - \widetilde{W}_{t-i}) \tag{30}$$

and in leading terms equation (27) can be written as

$$S_t = \theta(t) + \theta(t) \sum_{i=0}^{i_0} \frac{\vec{B}^* \cdot \vec{b}(t-1)}{1 - \widetilde{W}_{t-i}}$$
$$+ A\lambda \cdot \sum_{i=0}^{i_0} (\Delta_{\vec{z}} G_{t-1-i})_0 \tag{31}$$

or

$$S_t = \theta(t) + \sum_{i=0}^{i_0} \left( [\vec{C}_i^0 + \delta\vec{C}_i(t)] \cdot \vec{b}(t-i) \right)$$
$$+ A\lambda \cdot \sum_{i=0}^{i_0} (\Delta_{\vec{z}} G_{t-1-i})_0 \tag{32}$$

where

$$\theta(t) = S_{t-1-i_0} \prod_{i=0}^{i_0}(1-\tilde{W}_{t-i}) , \quad (33)$$

$$\vec{C}_i^0 \approx \theta(t)\vec{B}^* \quad (34)$$

and

$$\delta\vec{C}_i(t) \approx \theta(t)\vec{B}^*\tilde{W}_{t-i} . \quad (35)$$

The baseline $\theta(t)$ in (33) can be also presented in a form with trends as in (21). For example, for $\tilde{W}_t \ll 1$ and $t_0 \equiv t-1-i_0$

$$\theta(t) \approx S_{t_0} - S_{t_0}\sum_{i=0}^{i_0}\tilde{W}_{t-i} \approx S_{t_0}$$
$$+ (t-t_0)\frac{S_{t_0}}{i_0-1}\sum_{i=0}^{i_0-1}(\tilde{W}_{t-1-i} - \tilde{W}_{t-i}) \quad (36)$$

First two terms in (32) are in the agreement with generalized equation (24) for lags and random-coefficients mixed models. Mathematical expressions for coefficients (33) – (35) are approximations for the purpose of links demonstration only. Simultaneously, here we show the main difference between the mediaphysical approach and common statistical techniques: it is the last term in (32) which is absent in (24). This term is basically what this mediaphysical approach is about: the population connectivity in a distribution and the system "inertia". Observed results (catches) are not only direct responses to some external factors even with lags and coefficient fluctuations, but also consequences of time-dependent structures of the complex organized systems.

## 5. Model implementation and interpretation

Now we have a mathematical model to deal with. The starting point is an initial unit-density distribution, which has to be in agreement with market share or own-to-competitor ratio of average catches (sales or voters) from first available time points. If we do not have any other information about unit-density distribution, then we propose to use a truncated normal or "double" normal distribution in a manner to adopt known market share or the ratio above. If some parameters of the initial distribution are not defined from the very beginning, they can be fitted later to adjust system dynamics.

Then, we calculate system behavior during time in terms of Green's function, using described above equations, which for 1D system are (12) – (18) with the normalization (10). In Fig. 7 we present an example of population-distribution dynamics following this methodology and the corresponding time-dependent absorption (sales) on the own and competitor's attraction centers at $z=0$ and $z=1$.

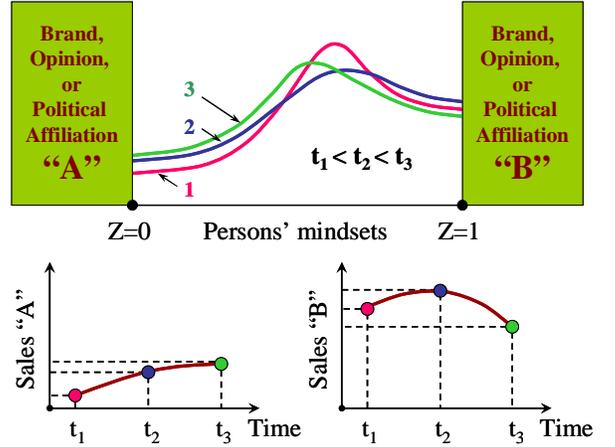

**Figure 7. Population-distribution dynamics and sales**

The dollar amount of sales (or number of buyers) is proportional to the Green's function at the absorption points or areas. In 1D case, own and competitor's sales are $S_0(t) = \lambda \cdot G(z,t)|_{z=0}$ and $S_C(t) = \lambda \cdot G(z,t)|_{z=1}$, respectively, with the single sales-to-Green's-function scaling factor $\lambda$.

In this approach, the majority of effects can be analyzed by numeric calculations and solutions of the equations. However, we can provide analytical estimations for some results: for instance, short-term effects (strictly speaking, next time point: while it can be next day, week or month) of advertising for a specific channel $i$. Thus, a relative increase in sale dollars per advertising dollar can be written as

$$\frac{1}{S_0}\frac{\partial S_0}{\partial b_{0i}} = \frac{\beta_0}{v_0+1} . \quad (37)$$

The similar equations are valid for both own and competitor's results. However, let us emphasize once more that it is valid for short-term effects. Long-term effects can be much more complicated and have to be analyzed under numeric computations.

Parameters of the model, which are not defined from the very beginning, have to be determined from the best fitting of model. Then, the adjusted parameters are to be used for forecasting, optimization of decisions, and for what-if analysis. To estimate these parameters we used an optimization procedure developed specially for this methodology (a kind of genetic-type algorithm with stochastic components). However, it is possible to use any reliable optimization procedure dealing with non-linear behavior with many local extremes and completely non-analytical definition.

The developed approach has been applied to a real media efficiency problem for a large company and demonstrated very good results despite the low initial correlations of factors and target variable.

## *6. Conclusions*

We developed a new *mediaphysical* approach to analyze a statistical dynamics of *real-life* systems, composed by many partly-unobserved units (usually human population), where mass communications (both internal and external) form their behavior. To apply this method to marketing, social life and politics, we introduced a concept of "persons' mindsets" space with distributed attractive and repulsive field-originated forces. We considered and formalized the concepts of the corresponding space dimensionalities, the distances in this space, and the relations between states of an individual mind and the distribution of all individuals. It is shown, that under reasonable assumptions, the system dynamics may be considered as a specific Markovian process, described in terms of Green's functions and Schrödinger-type equation, well known in theoretical physics. We showed links of the mediaphysical approach and such important concepts, as effects lasting in time (Box-Jenkins models, adstock in marketing), and random coefficients mixed models (or yield analysis), what makes mediaphysics naturally related with modern statistics. Simultaneously, it allows answering some questions, which are usually not or poorly addressed in traditional statistical and/or simulational sociophysical models:

1. A strong distinction between (a) data having mass structure with a distribution in some space, where only selected points can be observed (like marathon runners with TV camera maintained on start and finish lines only), and (b) data with only observed units (like single runner with a personal camera) was grounded. Respectively, for the first data type the mediaphysical approach is preferable. Traditionally, in statistics this distinction was not considered.
2. The proposed method organically allows considering both *short- and long-term effects* of marketing and propaganda in a consistence form, avoiding subjective definitions of lag or similar structures.
3. The technique allows formulating, checking and applying hypothesis about unobserved *population homogeneity or heterogeneity*, replacing traditional variables like "percent of educated women over time" by distributions of personal behaviors ("flexagility" or physical displacements). Thus, it strongly distinguishes variables, describing process and those describing population itself, which usually in statistics is bulked together.
4. The mediaphysical approach allows accounting for *interactions (influence) of the units between each other* (the leading topic of sociophysics, which is not usually addressed in statistics) in a more general way through self-consistent fields, that provides a unique opportunity to consider completely different components of social behavior together within one model, even when number of factors is big (unlike traditional sociophysics simulation with very limited number of factors).
5. It predicts *future system dynamics* from its history in such a way, that *system connectivity and inertia* are naturally taken into account. As a result, very important aspects of the forecasts (like prediction of hidden bumps of unobserved distributions) may be derived and used in practice.
6. For many problems the method allows replacing a simulation by analytical calculations or combining two approaches in a reasonable format.

Here, we just introduced basic ideas and methodology that we put in the proposed approach. These concepts should be developed much deeper. Among next-steps priorities we may list more detailed enhancements of the multifield scenarios and influential fields, non-trivial topologies of the considering space, maybe "tunnel transitions" (immediate jumps from one opinion to the opposite one), better optimization procedures, and many other problems. An important topic could be a what-if procedure for different scenarios and assumptions.

Authors are deeply grateful to David Hauser for stimulating discussions, constructive comments, and multifaceted undertakings that made this article possible. We thank Gary Gindler for critical reading of the text and making very useful comments.